\documentstyle[12pt]{article}
\renewcommand{\theequation}{\arabic{section}.\arabic{equation}}
\begin{document}

\topmargin=-1.0cm
\evensidemargin=0cm
\oddsidemargin=0cm

\newcommand{\BQ}{\begin{equation}}
\newcommand{\EQ}{\end{equation}}
\newcommand{\BQA}{\begin{eqnarray}}
\newcommand{\EQA}{\end{eqnarray}}
\newcommand{\half}{\frac{1}{2}}
\newcommand{\NN}{\nonumber \\}
\newcommand{\E}{{\rm e}}
\newcommand{\del}{\partial}
\renewcommand{\thefootnote}{\fnsymbol{footnote}}
\newcommand{\zm}[1]{\stackrel{\circ} {#1} }
\newcommand{\nzm}[1]{\widetilde {#1} }
\newcommand{\llangle}{\langle \langle}
\newcommand{\rrangle}{\rangle \rangle}

\date{}

\baselineskip=0.6cm

\renewcommand{\appendix}{\renewcommand{\thesection}
{\Alph{section}}\renewcommand{\theequation}{\Alph{section}.\arabic{equation}}
\setcounter{equation}{0}\setcounter{section}{0}}

\begin{titlepage}
\begin{flushright}
YITP-98-31\\
DPTN-9802\\
KOBE-TH-98-03 \\
{\tt hep-th/9806221}  
\end{flushright}

\begin{center}
{\LARGE Classical Aspects of the Abelian Higgs Model \\
on the Light Front }\\

\vskip1.0truein

{\large Kazunori Itakura}$^{(a)}$
\footnote{E-mail: {\tt itakura@yukawa.kyoto-u.ac.jp}},
{\large Shinji Maedan}$^{(b)}$
\footnote{E-mail: {\tt maedan@tokyo-ct.ac.jp}} and
{\large Motoi Tachibana}$^{(c)}$ 
\footnote{E-mail: {\tt tatibana@oct.phys.kobe-u.ac.jp}} 

\vskip0.2truein

\centerline{$^{(a)}$ {\it Yukawa Institute for Theoretical Physics,
Kyoto University, Kyoto 606-8502, Japan}}
\vspace*{2mm}
\centerline{$^{(b)}$ {\it Department of Physics, Tokyo National College
of Technology, Tokyo 193-8610, Japan}}
\vspace*{2mm}
\centerline{$^{(c)}$ {\it Graduate School of Science and Technology,
Kobe University, Kobe 657-8501, Japan}}

\end{center}
\vskip0.5truein \centerline{\bf Abstract} \vskip0.13truein

We investigate canonical structure of the 
Abelian Higgs model within the framework  of DLCQ.
Careful boundary analysis of differential equations, 
such as the Euler-Lagrange equations, leads us to a novel situation 
where the canonical structure changes in a drastic manner 
depending on whether the (light-front) spatial {\it Wilson line} 
is periodic or not. In the former case, 
the gauge-field ZM takes discrete values and
we obtain so-called ``Zero-Mode Constraints'' (ZMCs),
whose semiclassical solutions give a nonzero vev to the scalar fields.
Contrary, in the latter case, we have no ZMC and
the scalar ZMs remain dynamical as well as the gauge-field ZM. 
In order to give classically the nonzero vev to the scalar field,
we work in a background field which minimizes the light-front energy.

\end{titlepage}
\newpage
\baselineskip 20 pt
\pagestyle{plain}
\vskip0.2truein

\setcounter{equation}{0}

\vskip0.2truein

\section{Introduction}

Light-front (LF) quantization \cite{LF} and particularly
the method of Discretized Light-Cone Quantization 
(DLCQ) \cite{DLCQ} have a great impact not only upon nonperturbative study 
in field theories but also upon the M-theory as Matrix model in
string theory \cite{MT}.
A desirable property that the vacuum is simple or
trivial enables us to calculate 
mass spectra of bound states as well as their wave functions
using various nonperturbative techniques.
Such a simplification, however, is allowed only if we neglect the
{\it longitudinal zero momentum modes} (simply abbreviated as ZMs) 
of fields. This subtlety was first discussed seriously by  
Maskawa and Yamawaki \cite{MY} by setting the longitudinal direction
$x^-$ finite, which is the basic strategy of DLCQ,  
with periodic boundary condition.
The longitudinal momenta of fields are discretized and 
the ZMs can be safely treated.
They found that the ZMs of scalar fields are not dynamical and 
in fact subject to constraint relations called the 
{\it Zero-Mode Constraints} (ZMCs). 
After considerable study, it is now widely believed that 
nonperturbative treatment of the ZMCs is crucial for describing 
the spontaneous symmetry breaking (SSB) on the LF \cite{SSB}.

Compared to the scalar ZM, our understanding of the gauge-field ZM
is still insufficient though it might play an important role in
3+1 dimensional QCD. While its relevance for the topological structure 
such as the $\theta$-vacua, has been discussed in simple field 
theories \cite{theta}, more complicated situation  
where the topological effect coexists with SSB has never been investigated.
It can be best studied in the 1+1 dimensional Abelian Higgs model:
\BQA
&&{\cal L}= -\frac{1}{4}F_{\mu \nu}F^{\mu\nu} + 
            |D_{\mu}\phi|^2 -V(\phi),\label{lagrangian}\\
&&V(\phi) = \frac{\lambda}{4}\left({\phi}^*\phi - v^2\right)^2,
\label{fstreng,poten}
\EQA
where $D_{\mu} =\del_\mu -ieA_\mu$ is the covariant
derivative and $v^2 > 0$.
This is a very interesting model entangled with both ``Higgs mechanism'' 
and ``instanton'' associated to $\pi_1(S^1)={\bf Z}$.
Indeed the effect of instantons dramatically changes a naive 
perturbative picture: confinement of fractionally charged particles 
occurs even in broken phase \cite{AH}. 
So it is an intriguing question how the interplay of these two effects 
can be observed on the LF.

As the first step to attack this problem, 
we study in this letter classical aspects of the Abelian Higgs model 
{\it in general dimensions}. 
Within the Hamiltonian formulation, 
the appearance of the topological effect in 1+1 dimensions 
cannot be discussed until we quantize the system. 
Even at the classical level, however, we find a remarkable 
property of the model {\it in any dimension} which has never been discussed.

We work within a standard DLCQ method, i.e., the LF space 
$x^-=(x^0-x^1)/\sqrt2 $ is made compact  $x^- \in [-L, L]$ 
and we impose {\it periodic boundary conditions} for all fields.
The reason why we treat the periodic boundary conditions is that 
our standard knowledge in the LF formalism tells us that ZMs of 
scalar and gauge field are expected to be closely related to SSB 
and topological structure, respectively.

The content is as follows.
In the next section, we introduce the notion of the spatial Wilson
line using the Euler-Lagrange (EL) equation. 
The canonical structure of the model crucially depends on the periodicity
of the Wilson line. The periodic and non-periodic cases will be 
separately discussed in Secs. 3 and 4, respectively. 
The last section is devoted to discussions.

\setcounter{equation}{0}
\section{Dilemma in Canonical Structure}

Consider the EL equation for the Higgs field:
\BQ
D_+D_- \phi + D_-D_+\phi 
- D_iD_i\phi + \frac{\del V}{\del \phi^*}=0.\label{EL}
\EQ
The fastest way to derive ZMC in a (non-gauged) scalar theory is 
just integrate the EL equation over $x^-$ \cite{ROB}.
In our case, however, naive integration of (\ref{EL}) does not 
yield constraint relation because of the covariant derivative $D_-$.
In order to replace $D_-$ by the ordinary derivative $\del_-$, let us
introduce the (LF) spatial {\it Wilson line}  
\BQ
W(x^+,x^-,x_{\perp}) 
\equiv \exp \left\{ie\int_{-L}^{x^-}dy^-A_-(x^+,y^-,x_{\perp})\right\},
\EQ
where $x_{\perp}$ denotes the transverse directions. Then we have
\BQ
\del_-[2W^{-1}D_+\phi]
=W^{-1}\left[D_iD_i\phi +
ie\phi\Pi^--\frac{\del V}{\del\phi^*}\right],\label{EL2}
\EQ
where we used a formula $D_-f=W\del_-(W^{-1}f)$ and 
$\Pi^-=F_{+-}=\del_+A_--\del_-A_+$ is the conjugate momenta of $A_-$.
Note that the integration of the left-hand-side does not necessarily 
vanish because the Wilson line is not periodic in general.
Therefore we find that 
{\it only if the Wilson line is periodic} $W(-L)=W(L)$, the space
integration of (\ref{EL2}) generates a constraint analogous to the
usual ZMC:
\BQ
\int_{-L}^L dx^-\ W^{-1}
\left[D_iD_i\phi +ie\phi\Pi^--\frac{\del V}{\del\phi^*}\right] 
=0. \label{ZMC}
\EQ
A few comments are in order. First of all, 
there is no problem in regarding this equation as a constraint
though it includes the time derivative of $A_-$ through  $\Pi^-$.  
This is because (gauge-fixed) $A_-$ giving periodic Wilson line 
is restricted to discrete values. 
More details are in the next section.

Next, this somewhat heuristic observation suggests that 
the canonical structure itself depends on the periodicity of the Wilson line.
In fact, the necessity to distinguish these two cases is required when we
try to determine Lagrange multipliers in the usual canonical procedure.
The consistency condition of a Lagrange multiplier $\lambda$ 
for a (primary) constraint $\theta=\Pi_{\phi^*}-D_-\phi$ is 
\BQ
D_-\lambda=\half \left(D_iD_i\phi +ie\phi\Pi^--\frac{\del V}{\del
\phi^*}\right)\equiv K.\label{PDE}
\EQ
Ambiguity in the Lagrange multiplier is intimately 
related to a new constraint. Therefore, it is essential to obtain a 
general solution to this differential equation. 
Indeed, the counterpart in the usual 
scalar theory is just a differential equation 
$2\del_-\lambda=\del_i\del_i\varphi-\del V/\del \varphi$ 
and its general solution 
subject to $\lambda(-L)=\lambda(L)$ allows a ``zero mode'' to be
undetermined, which implies the existence of the ZMC \cite{MY}. 
The general solution to eq.(\ref{PDE}) without boundary consideration 
for $\lambda$ is given by 
\BQ
\lambda(x)= W(x)\int_{-L}^{x^-}dy^- W^{-1}(y)K(y) + C\ W(x),\label{gene_sol}
\EQ
where $C$ is an integral constant. The second term comes from 
a homogeneous equation $D_-\lambda=0$. Imposing further 
$\lambda(-L)=\lambda(L)$, a natural requirement from the periodicity 
of the Hamiltonian, we must distinguish two cases depending on the
periodicity of $W(x)$.
In the case where $W(x)$ is {\it not periodic}, i.e., $W(L) \neq W(-L)$,
we can completely determine $\lambda$ from the requirement 
$\lambda(-L)=\lambda(L)$. So there is no residual constraint in this
case and therefore 
{\it both the scalar and gauge ZMs are dynamical}. 
On the other hand, in the case of the periodic Wilson line, 
i.e., $W(L) =W(-L)= 1$, the integral constant $C$ leaves undetermined and
we have the following {\it secondary constraint}
\BQ
\int_{-L}^L dx^-\ W^{-1}(x)K(x)=0.
\EQ
This is the same as eq.(\ref{ZMC}). 
The canonical structure of the model changes drastically depending on
the periodicity of the Wilson line.
As is evident from these arguments,
this somewhat strange situation is directly related to a problem 
whether we can define the inverse of the differential operator $D_-$ 
and, in other words,  whether $D_-$ has zero eigen-value or not.

The final comment is on physics.
 From our conventional knowledge in the LF formalism, 
we naively expect that ZMC for scalars may be necessary for 
describing the Higgs mechanism whereas the (dynamical) gauge-field ZM 
will be responsible for nontrivial vacuum structure in 1+1 dimensions. 
Nevertheless, as we saw, ZMC does not necessarily exist 
and it rather emerges in a very special case with measure zero 
in the phase space. 
When the ZMC exists, the gauge-field ZM is restricted to discrete values.
On the other hand, the more general case allows gauge-field ZM to be fully
dynamical but there is no ZMC and now we cannot follow the conventional 
discussion to describe the symmetry breaking.   
So we are in a very dilemmatic situation.

In the following sections, we discuss these two cases separately 
and restrict our consideration to the 1+1 dimensional case for brevity.
We can do that without loss of generality.

\setcounter{equation}{0}
\section{Periodic Wilson line and the Zero-Mode Constraint}

Let us discuss an appropriate decomposition of the scalar field.
The existence of ZMC does not necessarily imply that the naive 
decomposition of scalars into zero and non-zero modes is a good one. 
To see this, we first perform the gauge fixing.
Since the Light-Cone axial gauge in compact space inevitably misses
the ZM of $A_-$, i.e., $\zm{A}_- = (1/2L)\int_{-L}^L 
dx^- A_-(x)$ remains unfixed, 
zero-mode separation of $A_-$ has a physics meaning. 
 After gauge fixing, requiring periodicity to the Wilson line 
restricts $\zm{A}_-$ to be discrete value 
\BQ
\zm{A}_-=\frac{\pi n}{eL},\qquad n\in {\bf Z}.
\EQ
Since the nonZM $\widetilde{A}_-$ is fixed to be zero  
in the LC gauge and $\del_+\zm{A}_-=0$, 
eq.(\ref{ZMC}) does not include time derivative and thus can be 
understood as a constraint relation.\footnote{More strictly, you can 
impose $W(L)=1$ as a constraint in this case. Then, from the
consistency condition of the constraint, 
the secondary constraint $\zm{\Pi}\!{}^-=\del_+\zm{A}_-=0$ is derived.} 
Now the periodic Wilson line 
becomes $W(x)=e^{i\pi n (x^-+L)/L}$. This means that 
the ambiguous mode of the Lagrange multiplier is not 
a zero mode but a mode with frequency $\pi n /L$ (see
eq.(\ref{gene_sol})), and also that
a scalar field with frequency $\pi n /L$ should be taken  
as a constrained variable.
It will be convenient to redefine the scalar field 
so that the zero mode of a new variable becomes constrained:
\BQ
\phi_n(x) \equiv e^{-i\frac{\pi n}{L}x^-}\phi(x).
\EQ
New field also satisfies the periodic boundary condition and this
change of variable is essentially equivalent to the 
{\it large gauge transformation}.
The ZMC (\ref{ZMC}) is rewritten in terms of $\phi_n$ as
\BQ
\int_{-L}^L dx^-\ \left(ie\phi_n\nzm{\Pi}^- - 
\frac{\partial V}{\partial \phi_{n}^*}\right)=0.
\label{ZMC2}
\EQ
Explicit form of $\nzm{\Pi}^-$  is given by solving the Gauss law 
$\del_-\Pi^-=ie [(D_-\phi)^*\phi - \phi^*D_-\phi]$, 
\BQ
\nzm{\Pi}^-=\int_{-L}^L dy^- \ \nzm{\theta}(x^--y^-) 
ie (\del_-\phi_n^*\phi_n-\phi_n^*\del_-\phi_n),
\EQ
where $\nzm{\theta}(x^--y^-)$ is the periodic step function $\del_-^x\nzm{\theta}(x^--y^-)=\delta(x^--y^-)-1/2L $.
Equation (\ref{ZMC2}) can be understood as a constraint relation 
for the zero mode of the new variable and the decomposition 
$\phi_n =\ \zm{\phi}_n+\ \widetilde{\phi}_n$ becomes useful.
That is why we called eq.(\ref{ZMC2}) the ZMC from the beginning.
Now all the ZMs ( $\zm{\phi}_n, \ \zm{\phi}_n\!\!\!{}^*$ and $\zm{A}_-$
) can be treated as non-dynamical. Following the standard procedure, 
we obtain the Dirac brackets between physical variables:
\BQA
&&\left\{\nzm{\phi}_n(x),\ \nzm{\phi}_n^*(y) \right\}_{\rm DB}
= -\frac{1}{4}\ \epsilon(x^- - y^-), \\
&&\left\{\nzm{\phi}_n(x),\ \nzm{\phi}_n(y)\right\}_{\rm DB}
= \left\{\nzm{\phi}_n^*(x),\ \nzm{\phi}_n^*(y)\right\}_{\rm DB}=0.
\EQA
We can easily go to quantum theory by replacing these by commutators.
Let us evaluate eq.(\ref{ZMC2}) in a semiclassical treatment 
($\hbar$-expansion). 
In the lowest order, any ZM should not have operator part.
Since operator contribution will come from the nonzero modes,
we here assume that the lowest (classical) ZM  does not depend 
on the nonzero mode $\nzm{\phi}_n$. So in this simple approximation, 
eq.(\ref{ZMC2}) becomes just
\BQ
\left(\zm{\Phi}_n\!\!\!{}^* \zm{\Phi}_n - v^2 \right)\zm{\Phi}_n = 0,
\EQ
and the solution is
\BQ
\zm{\Phi}_n = 0, \quad v\ e^{i\alpha},
\EQ
where $\zm{\Phi}_n$ denotes the classical part of $\zm{\phi}_n$ and
$\alpha$ is an arbitrary constant.
Rewritten in terms of the original variable, 
the solution $\zm{\Phi}_n = v e^{i\alpha}$ corresponds to 
nonzero mode.
Nevertheless, it can be made into a zero mode by the large
gauge transformation and thus gives nonzero vacuum expectation value.
This is equivalent to evaluating the ZMC with $n=0$ from the beginning. 
In addition, the Hamiltonian has no $n$-dependence after we insert the 
solution $\zm{\Phi}_n$.

\setcounter{equation}{0}
\section{Non-Periodic Wilson line}

In this case, there is no ZMC and thus all the ZMs should be taken as 
dynamical variables. This means that we cannot utilize the method 
of solving ZMC to describe SSB.
The situation here is rather similar to the usual equal-time 
calculation: we do not have to have recourse to such a special method 
restricted to the LF formalism.
Since this model shows SSB in the tree (classical) level, 
we should be able to describe it in a classical treatment.
This is possible if we proceed analogously to the 
{\it background field method}.
Let us first look for a configuration (in the LC axial gauge) 
which minimizes the LF energy 
\BQ
P^-= \int_{-L}^{L}dx^{-}\left[ \frac{1}{2}(\Pi^-)^2 + V(\phi) \right]
\ge 0.
\label{ham}
\EQ
Note that we further need to impose the Gauss law,
unlike the equal-time calculation.
The LF energy becomes zero if and only
if field configuration is given by
\BQA
\zm{A}_-
&=& \frac{\pi N}{eL}, \qquad N \in {\bf Z},\\
\phi &=& v\ e^{i(\frac{\pi N}{L}x^-+\alpha)},
\EQA
where $\alpha$ is an arbitrary constant and can be set to be zero.
This configuration is exactly the same as the classical solution of 
eq.(\ref{ZMC}) in the periodic Wilson line case.
Therefore we can give nonzero vev to the scalar field
even {\it without ZMC} if we expand the fields around this
classical energy-minimized configuration.
Note also that this configuration gives $P^+=0$ and therefore is
equivalent to the configuration giving zero equal-time energy.

We consider the canonical structure in the energy-minimizing 
background field.
Let us introduce $V_\pm$ and $\varphi_N$ by
\BQA
&&A_+=V_+\ ,\\ 
&&A_-=\frac{\pi N}{eL} + V_-\ ,\\
&&\phi_N=v+\varphi_N\ .
\EQA
We again introduced $\phi_N=e^{-i\frac{\pi N}{L}x^-}\phi$ for convenience. 
Then we decompose fields into zero and nonzero modes $
V_\pm=\zm{V}_\pm+\ \nzm{V}_\pm $ and $\varphi_N=\ 
\zm{\varphi}_N+\ \nzm{\varphi}_N$.
To avoid the singularity $\zm{V}_-=0$ which gives the periodic Wilson
line, it should be 
$
0 < \ \zm{V}_- < \pi/eL.
$
This restriction with a given $N$ means that we work in a fixed 
sector with respect to the large gauge transformation.

After lengthy but straightforward calculations, the Dirac brackets
between dynamical fields are obtained (in the LC axial gauge:
$\zm{V}_+=0,\ \nzm{V}_-=0$ and $\nzm{\Pi}^-+\del_-\nzm{V}_+=0$) 
as follows:
\BQA
\Big\{\zm{V}_-,\ \zm{\Pi}\!\!{}^-\Big\}_{\rm DB} &=& \frac{1}{2L}, \nonumber \\
\Big\{\zm{\varphi}_N ,\ \zm{\varphi}_N\!\!{}^* \Big\}_{\rm DB}
&=& \frac{1}{2L}\frac{1}{2ie\zm{V}_-}, \nonumber \\
\Big\{\zm{\Pi}\!\!{}^-,\ \zm{\varphi}_N \Big\}_{\rm DB}
&=& \frac{1}{4L\zm{V}_-}(v+\zm{\varphi}_N), \\
\Big\{\zm{\Pi}\!\!{}^-,\ \nzm{\varphi}_N(x)\Big\}_{\rm DB}
&=& \frac{1}{2L}\int_{-L}^Ldy^-\ 
ie \nzm{\varphi}_N(y)\ \frac12 G^*(y,x), \nonumber \\
\Big\{\nzm{\varphi}_N(x),\ \nzm{\varphi}_N^*(y)\Big\}_{\rm DB}
&=&  -\frac12\ G(x,y).  \nonumber
\EQA
where $\zm{\Pi}\!\!{}^-=\del_+\zm{V}_-$ is a conjugate momenta of
$\zm{V}_-$ and 
$G(x,y)$ is Green's function defined through 
$
(\del_-^x-ie\zm{V}_-)G(x,y) = \delta(x^--y^-) - 1/2L\ .
$
 From this, it is evident that the number of ZM degrees of freedom 
is two ($\zm{V}_-$ and $\zm{\varphi}_N$). In quantum theory, 
these ZMs should be responsible for a nontrivial vacuum structure 
such as the $\theta$-vacua \cite{IMT}.

\setcounter{equation}{0}
\section{Discussions}

In this letter
we introduced a notion of the periodicity of the Wilson line 
and found that the canonical structure of the ZM sector changes 
drastically depending on its periodicity. In almost all the phase
space except countable points,
both the scalar and gauge-field ZM are dynamical. 
At the exceptional points, the gauge-field ZM takes discrete values 
and the scalar ZM becomes constrained. We should mention here that
this kind of canonical structure has been already
pointed out in the 3+1 dimensional $SU(2)$ Yang-Mills theory \cite{Franke}.
At the points where $\zm{A}_-\!\!\!{}^a=\delta^{a3} n\pi/gL$, 
there exist two constraint relations:
\BQ
\int_{-L}^{L}dx^-e^{\frac{i\pi n}{L}x^-}(G^1+iG^2)=0,\label{YM}
\EQ
where $G^a = (D_iF_{-i})^a$ and $(D_iF_{+-}+D_jF_{ji})^a$ $(a=1,2)$.
These are also originated from the zero eigenvalue problem of the
covariant derivative $D_-$. Therefore the same kind of constraint will 
exist in any gauge theory.
In Ref. \cite{Franke}, quantum theory in the case with 
constraints (\ref{YM}) was not developed due to its complexity and 
also the physics consequences of these constraints are still not clear.
In our model, however, the ZMC has a significant meaning that 
we could give a nonzero vev to the scalar by solving it semiclassically.
We expect that the constraints (\ref{YM}) also will give some 
nontrivial effects on the theory.

One of the most important question is whether the physics in
these two cases, which are canonically distinct from each other, 
is continuously connected.
Its complete solution is not yet given at present.
However, as for the symmetry breaking in this model, 
it should be discussed even in the classical theory.
So we managed to construct a classically breaking theory for both cases. 
In the general case (nonperiodic Wilson line), we could express the
symmetry breaking by expanding fields around energy-minimizing 
configuration. This procedure may correspond to evaluating the
effective potential {\it classically}. With dynamical zero modes 
in this case, we will be able to evaluate the vacuum energy as 
in the equal-time calculation. It should be checked after quantization  
whether our background field is consistent or not.

One more necessary ingredient, the $\theta$-vacua, should 
also be discussed after quantization. 
This is because in the Hamiltonian formulation, 
topological effects such as instantons will be observed as
the ``quantum tunneling'' between multiple vacua.
The multiplicity of the vacuum is generated by the large gauge 
transformation which is a displacement symmetry for the gauge-field ZM. 
We expect that thorough treatment of this symmetry 
with appropriate continuum limit $L\rightarrow \infty$ will
lead to a separation of the 1+1 dimensional model from 
the higher dimensional models, 
where there should be no topological effects.

All these will be discussed in the future work \cite{IMT}.

\vskip0.3truein
\centerline{{\it ACKNOWLEDGMENTS}}
The author M.T. would like to thank to our colleagues
in Kobe University for valuable comments and discussions. 
We are especially grateful to K. Harada, T. Matsuki, M. Taniguchi 
and K. Yamawaki for their valuable comments and various discussions.


\newpage

\end{document}